\documentclass[11pt, a4paper]{article}
\usepackage{moriond,epsfig}
\def\be{\begin{equation}}
\def\ee{\end{equation}}
\def\bea{\begin{eqnarray}}
\def\eea{\end{eqnarray}}

\begin{document}
\vspace*{4cm}
\title{The Gould Belt, star formation, and the local interstellar medium}

\author{Isabelle A. Grenier}
\address{AIM, Service d'Astrophysique, CEA Saclay, 91191 Gif/Yvette, France}

\maketitle\abstracts{The history of the local medium, within a few hundred parsecs, is dominated by the
evolution of the Gould Belt. The event that triggered this star-forming region and molded the gas
distribution is still unknown. Its orientation and extent are reasonably well determined and its expansion
matches the space and velocity distributions of many large HI and H$_2$ clouds within half a kiloparsec. The
present rim coincides with most of the nearby OB associations, but their mean velocity does not seem to be
related to the Belt expansion. The Belt age is uncertain by a factor of 2 because of the discrepancy found
between the dynamical timescale (20 to 30 Myr) and the stellar ages (30 to 60 Myr). The stellar content is
derived from kinematic studies for massive stars and from X-ray observations for young solar-mass ones.
Whether star formation is active along the rim or spread over a larger fraction of the disc is debated. The
Belt flatness and its tilt remain very difficult to interpret. Various scenarii involve the impact of a
high-velocity cloud, a cascade of supernovae, the dissolution of a rotating system, or the braking of a
supercloud entering the spiral arm. Because of the enhanced star formation, the Belt supernova rate over the
past few million years has been 3 to 4 times larger than the local Galactic rate. The corresponding pulsars
may be responsible for the population of unknown $\gamma$-ray sources associated with the Belt. The higher
rate also implies an enhanced cosmic-ray production locally.}

\noindent {\small¥{\it Keywords}: Galaxy: solar neighbourhood, stars: early-type, stars: supernovae, stars:
pulsars, ISM: kinematics and dynamics, Gamma rays: observations}

\section{The Belt geometry}
Many stars have formed locally over the past few $10^7$ years in a surprisingly flat and inclined disc named
the Gould Belt. The Sun happens to be crossing this structure. The asymmetry about the Galactic plane of the
the bright-star distribution was first pointed out by Sir John Herschel \cite{herschel_1847} in 1847 and
studied in 1874 by Benjamin A. Gould \cite{gould_1874} who determined the Belt orientation with respect to
the Galactic plane. From the decomposition of the spatial distributions of the two intersecting discs in the
\textit{Hipparcos} data \cite{torra00a}, Torra et al. found that 60 to 66\% of the massive stars with ages
$<$ 60 Myr and distances $<$ 600 pc belong to the Belt. This fraction significantly decreases to 42-44\% for
ages between 60 and 90 Myr. In terms of spectral types, 44\% and 36\% of the O-B2.5 and O-B9.5 stars within 1
kpc are respectively linked to the Belt \cite{cabrera99}.

Significant changes in the Oort constants with stellar age within 600 pc reflect the Belt kinematical
influence. A nearly pure differential Galactic rotation is found for the old stars ($>$ 90 Myr) whereas the
younger ones exhibit a marked decrease in A and a large negative B value that suggests that the whole system
is rotating \cite{lindblad97,torra00a}. A positive K term, indicative of expansion, is measured for stars
younger than 30 Myr. These peculiarities remain after removing the Sco-Cen and Ori OB1 associations from the
sample, so stars outside these famous clusters along the Belt rim participate to the unusual kinematics. The
true velocity field may, however, be more complex than the Oort constants would suggest. There is a
systematic gradient in the vertical velocities of the young stars along the Galactic plane \cite{comeron99}
and the ascending node longitude for vertical oscillation is $337^{\circ} \pm 20^{\circ}$ instead of
$296^{\circ} \pm 2^{\circ}$ for the Belt orientation \cite{perrot03}.

Whereas the massive-star content has been extensively studied, less is known about the low-mass star
production. Young (30--80 Myr old) Lithium-rich solar-mass stars show up as X-ray sources because of their
active coronae. Even though the X-ray horizon is limited by interstellar absorption to 150-300 pc for sources
with luminosities of $(0.3-3) 10^{23}$ W typical of young late-type stars, such stellar sources in the ROSAT
All-Sky Survey nicely trace the Belt in the sky \cite{guillout98b}.

The Gould Belt also contains interstellar clouds and has been early associated with an expanding HI ring
\cite{lindblad67}. The fact that dark clouds participate to the expansion was recognized 20 years later
\cite{taylor87}. Famous H$_{2}$ complexes, such as Orion, Ophiuchus, and Lupus, have long been related to the
Belt, but more recently mapped complexes, such as Aquila Rift, Cepheus, Cassiopeia, Perseus, and Vela appear
to be part of the expanding shell as well \cite{perrot03}.

\begin{figure}
\centerline{\includegraphics[width=0.5\linewidth]{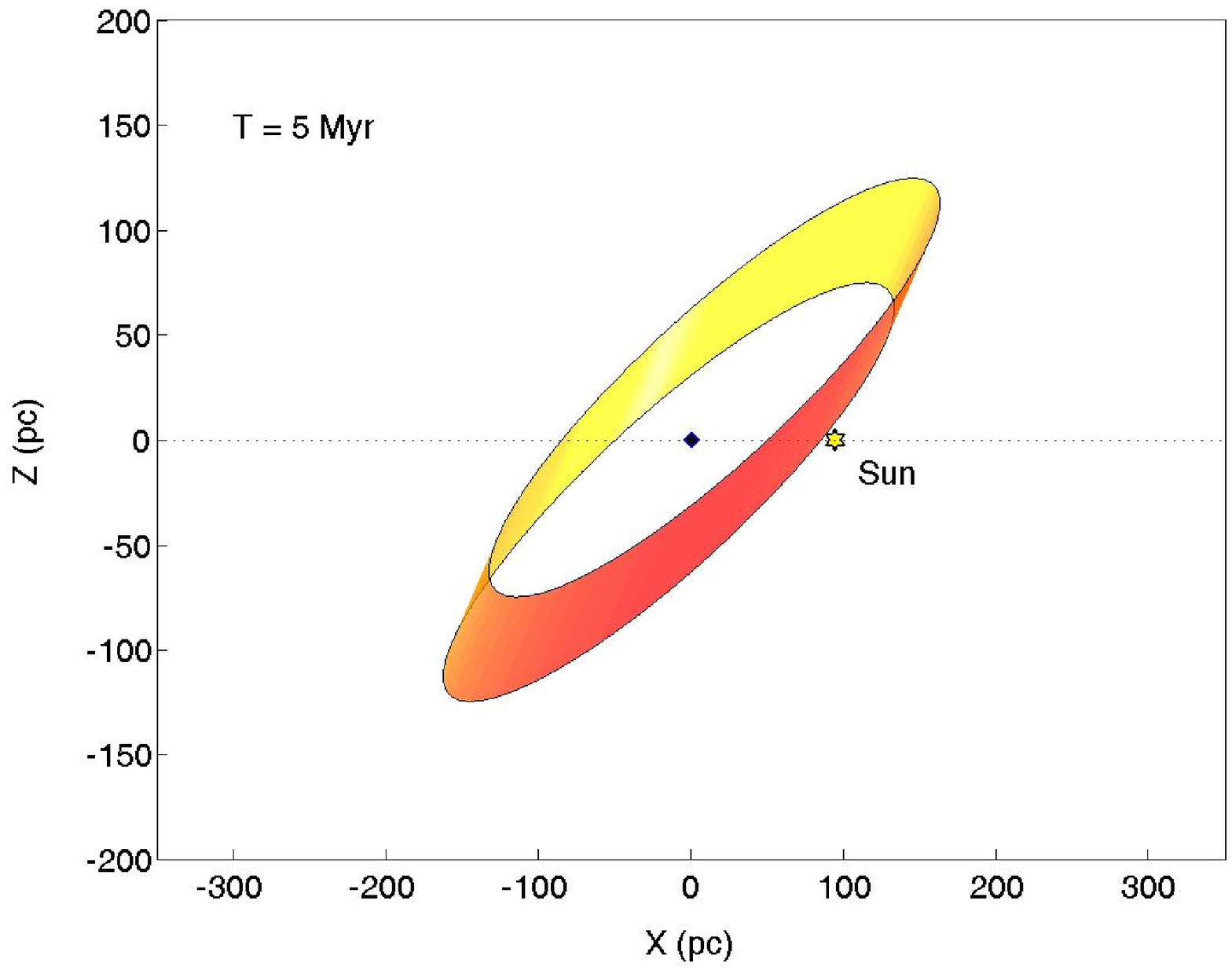}
\includegraphics[width=0.5\linewidth]{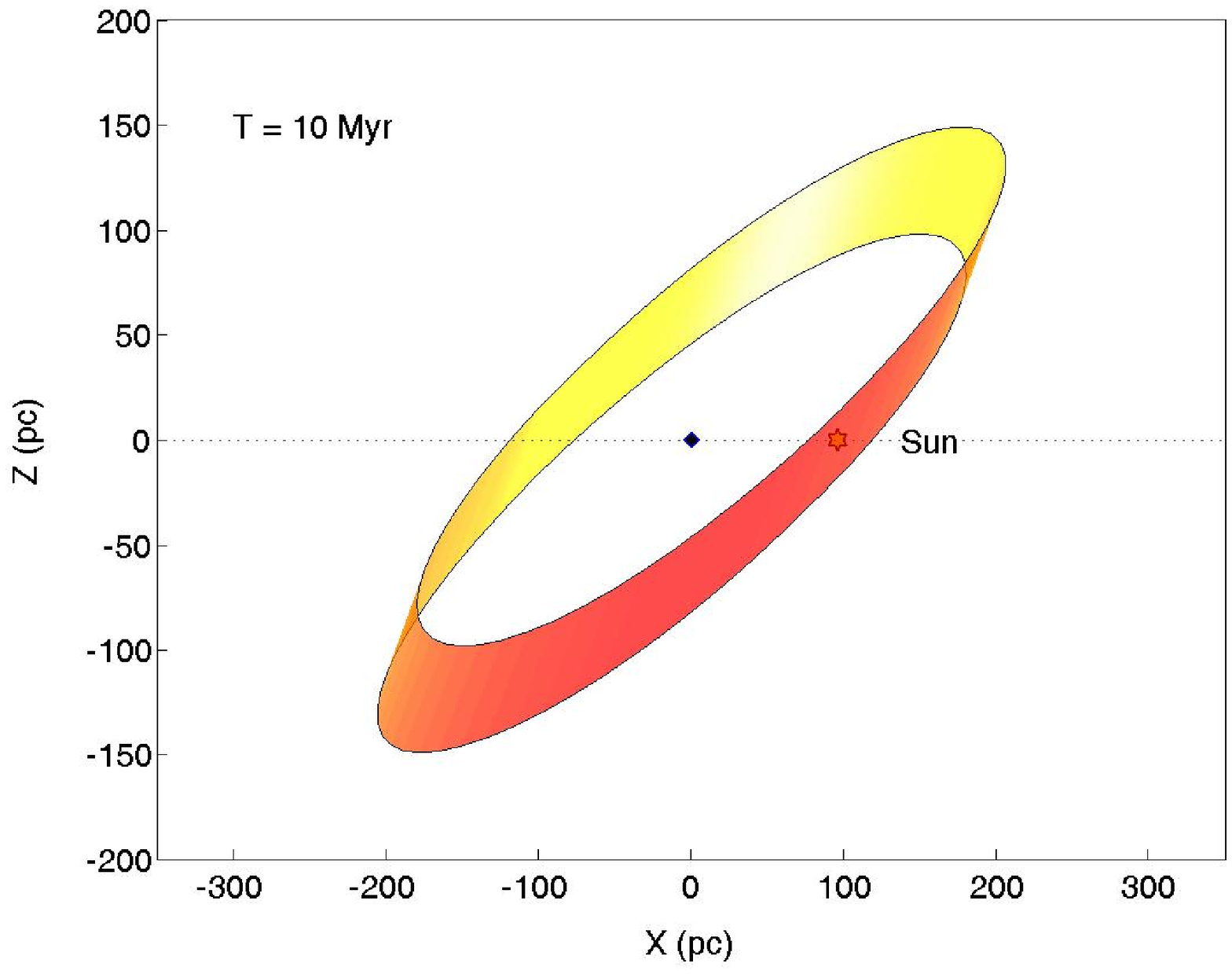}}
\centerline{\includegraphics[width=0.5\linewidth]{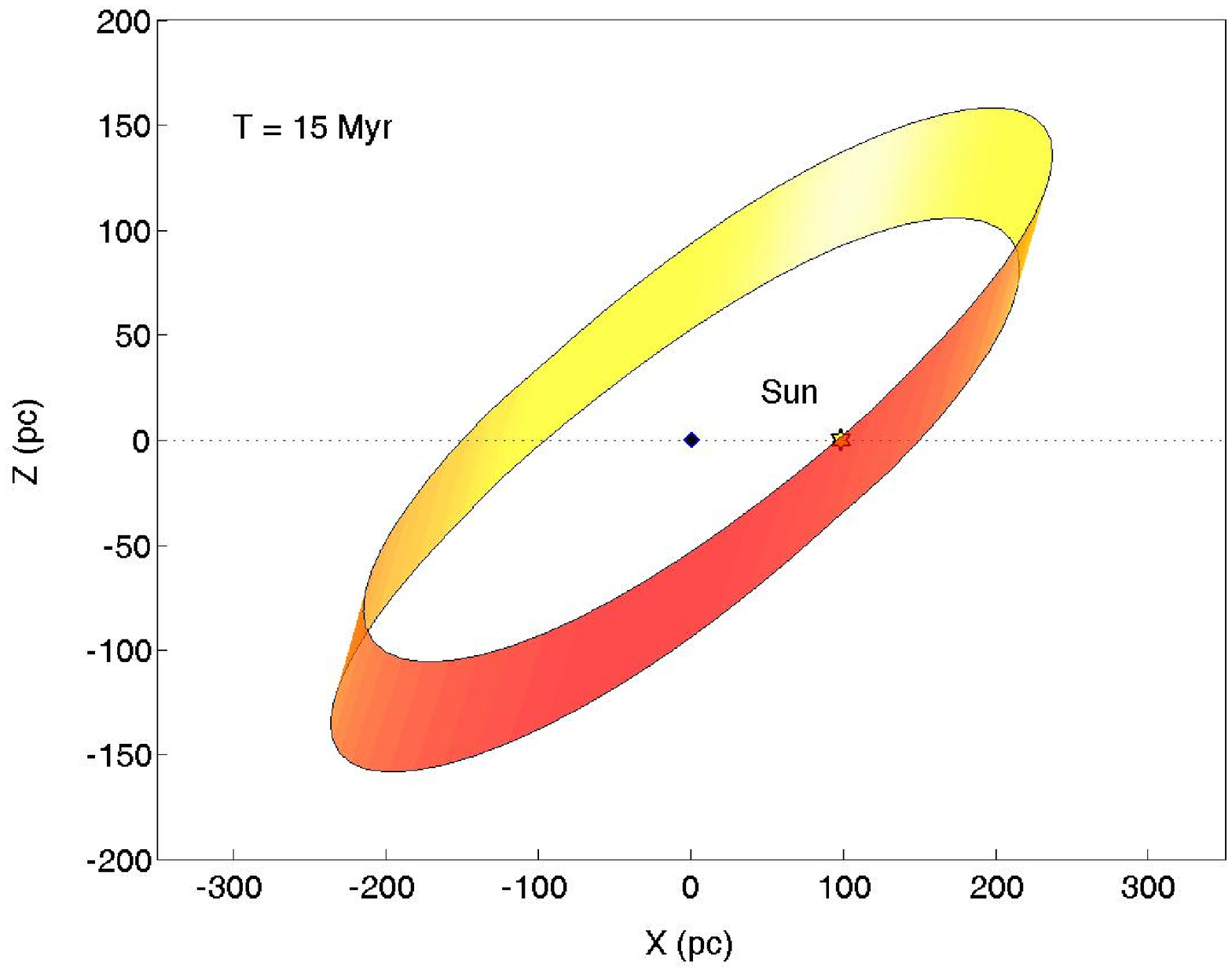}
\includegraphics[width=0.5\linewidth]{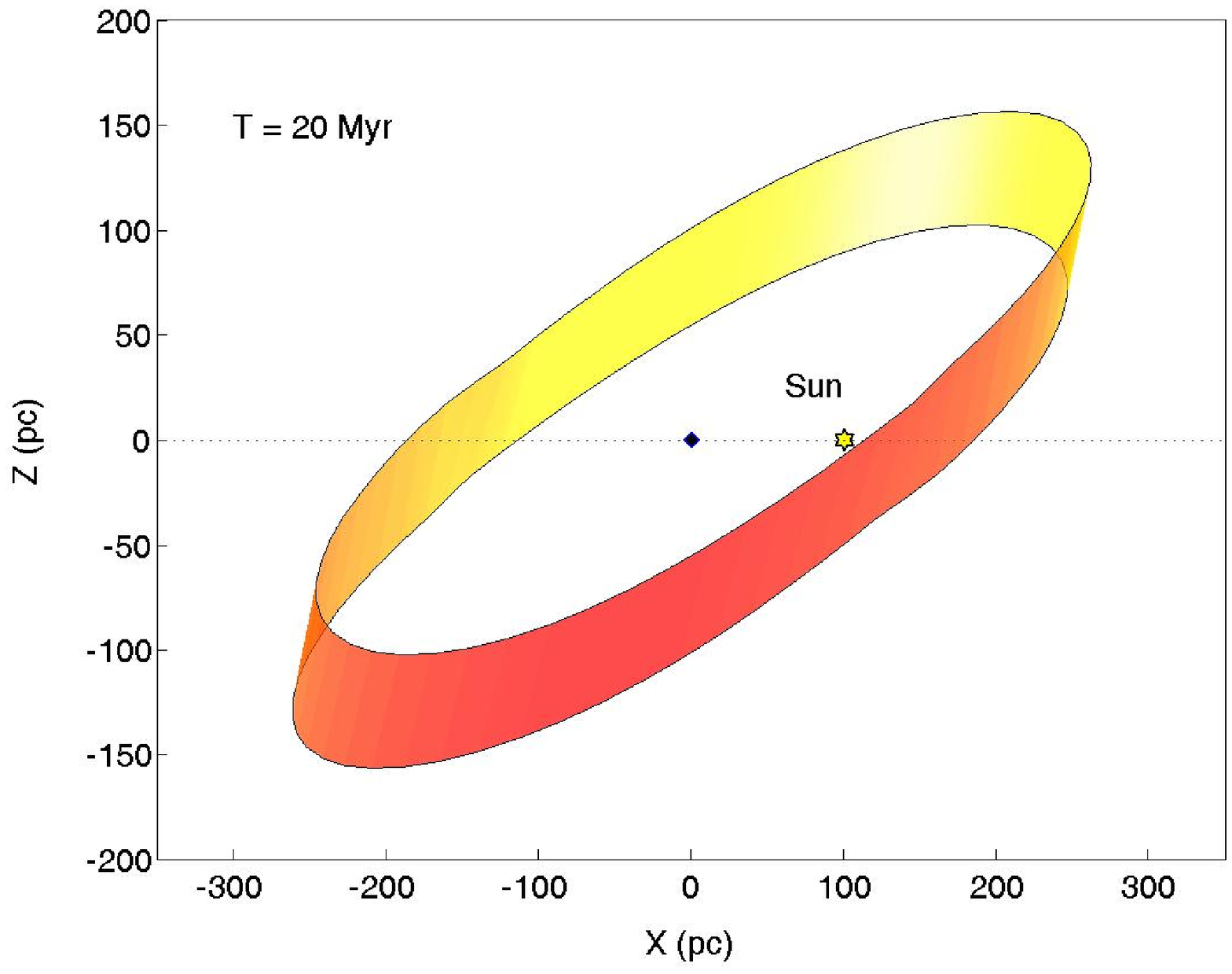}}
\centerline{\includegraphics[width=0.5\linewidth]{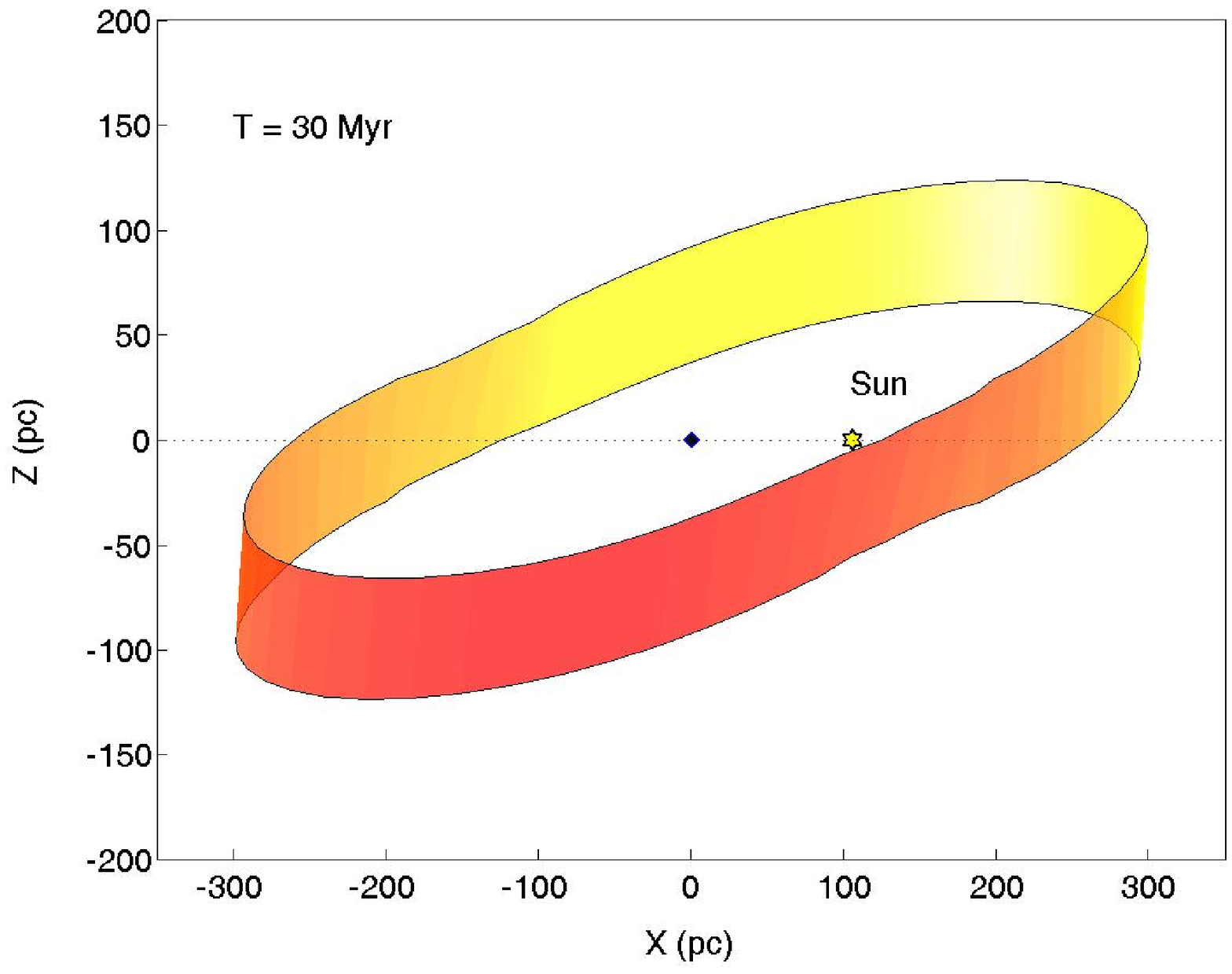}
\includegraphics[width=0.5\linewidth]{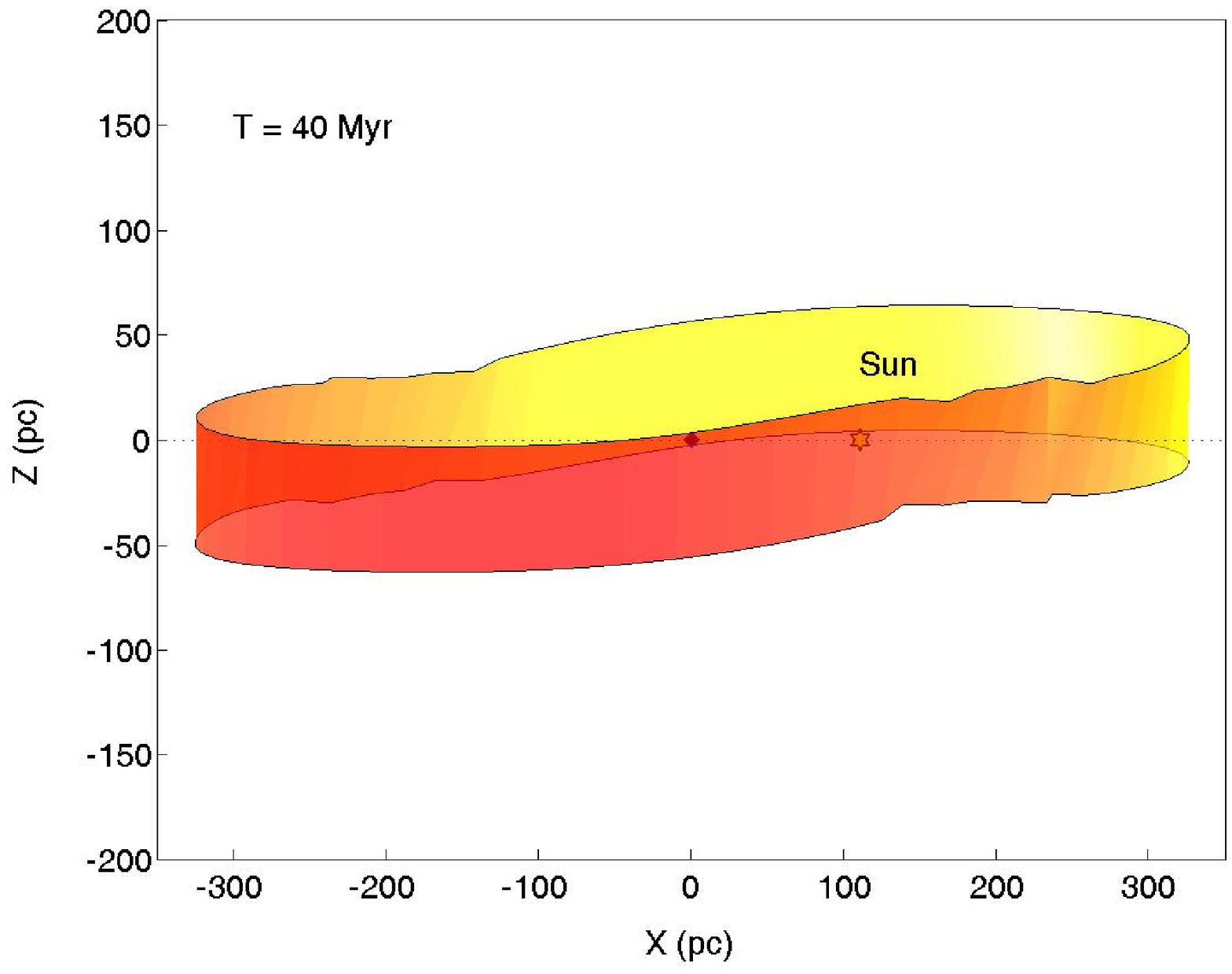}}
\caption{The Gould Belt evolution as seen at different epochs after the outburst, in a plane perpendicular to
the Galactic plane, centred on the Belt centre. The x axis points to the Galactic centre and the location of
the Sun, nearly half way to the rim, is marked by an asterisk.}\label{fig:Beltmovie}
\end{figure}

The present size of the gas shell was first estimated by comparing the radial velocities in the Lindblad HI
ring with the 2D expansion of a shock wave inside the Galactic plane \cite{olano82}, then with the 3D
expansion of a superbubble in a uniform medium \cite{moreno99}. The dynamical evolution has been revisited to
allow 3D expansion in a non uniform medium and to compare with the location and motion of all the nearby HI
and H$_2$ clouds \cite{perrot03}. An inclined cylindrical shock wave, with thickness $H$, has been used. It
sweeps momentum from the ambient medium where the gas density varies with altitude above the Galactic plane
according to the 90, 225, and 400 pc scale heights that describe the HI layer \cite{dickey90}, and to the 74
pc scale height of the local CO gas \cite{dame87}. Because of the Galactic differential rotation, the
circular section of the Belt rapidly evolves into an elliptical one which precesses with time. The rim
slightly takes an hourglass shape from its faster expansion in the more rarefied medium at high altitude. The
Belt further warps and falls back in the Galactic gravitational potential. A density gradient with
Galactocentric distance has little effect on the evolution. The sequence that best matches the present
location and longitude-latitude-velocity distribution of all the nearby HI and CO clouds at $|b| > 5^{\circ}$
is displayed in Figure \ref{fig:Beltmovie}. The current Belt geometry is close to that depicted in the 30 Myr
plot and the last plot illustrates the Belt in 10-15 Myr from now.

The dimensions \cite{perrot03} that best fit the cloud data are a height H of 60 pc and semi-major axes $a =
354 \pm 5$ pc and $b = 232 \pm 5$ pc, in good agreement with the sizes derived from the sole HI data for the
expansion of a 2D ring \cite{olano82} ($360 \times 210$ pc) or a 3D superbubble in a uniform medium
\cite{moreno99} ($341 \times 267$ pc). The present inclination of $17.2^{\circ} \pm 0.3^{\circ}$ to the
Galactic plane nicely compares with the latest stellar estimates ($16^{\circ}-22^{\circ}$ for massive stars
younger than 60 Myr \cite{torra00a} and $17.5^{\circ}-18.3^{\circ}$ for O-B stars \cite{cabrera99}). The
larger inclination of $27.5^{\circ} \pm 1^{\circ}$ indicated by the young solar-mass stars is biased by the
dominant Sco-Cen associations within the X-ray visibility horizon \cite{guillout98b}.  The current Belt
centre is found at $104 \pm 4$ pc from the Sun, toward $l = 180^{\circ} \pm 2^{\circ}$, so the Sun is nearly
half way to the rim. The ascending node longitude $l_{\Omega} = 296.1^{\circ} \pm 2.0^{\circ}$ is
$10^{\circ}$ higher than the values obtained from the massive and young low-mass stars
\cite{torra00a,guillout98b}, possibly reflecting the time-lag between stellar birth and the slowly precessing
rim.

\begin{figure}
\psfig{figure=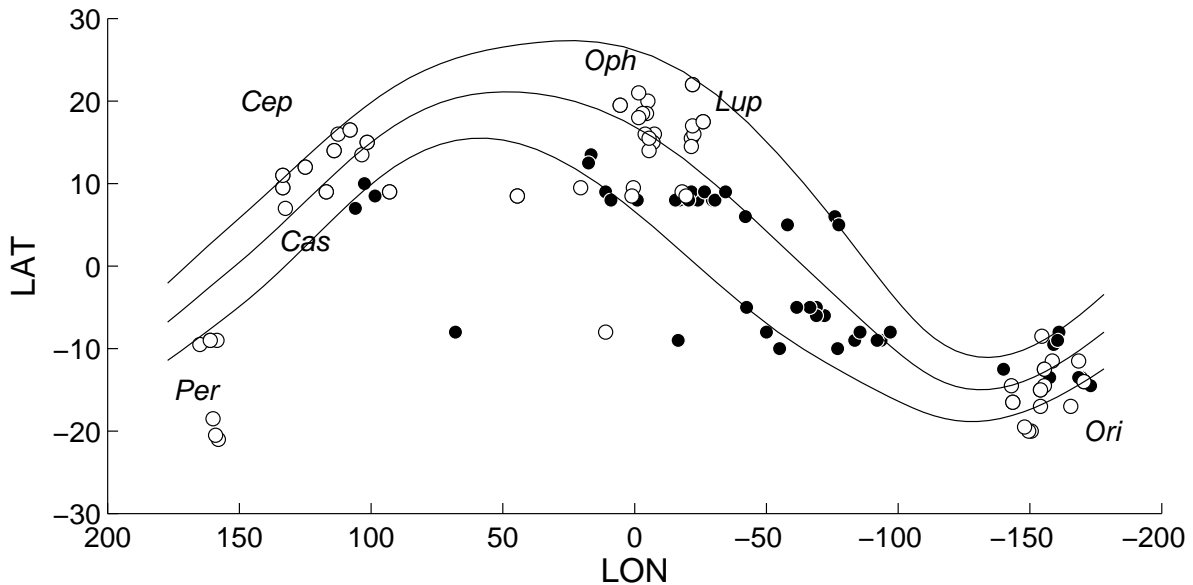,height=8cm} \caption{Galactic longitude and latitude distribution of the nearby HI
and H$_{2}$ clouds at $|b| > 5^{\circ}$ and the trace of the 60-pc-thick disc of the Belt across the sky. The
Belt angular extent is larger near Ophiuchus where the rim is closer to the Sun.} \label{fig:Beltlb}
\end{figure}

The Belt position and orientation differ from previous estimates because of the use of \textit{Hipparcos}
distance information and of major H$_{2}$ complexes in the second quadrant. They had not been firmly
associated with the Belt before, but their direction, distance, and velocity appear to be quite consistent
with the modelled Belt and with the HI Lindblad ring \cite{perrot03}. In fact, nearly all the local H$_2$
complexes at $|b| > 5^{\circ}$ seem to participate to the shell, except the very nearby Taurus and R CrA
clouds that are well inside the Belt, and the Chamaeleon clouds that belong to the Local Arm. Figure
\ref{fig:Beltlb} shows the Belt trace across the sky with respect to the clouds.

\section{Star-formation in the Belt}
\begin{figure}
\centerline{\psfig{figure=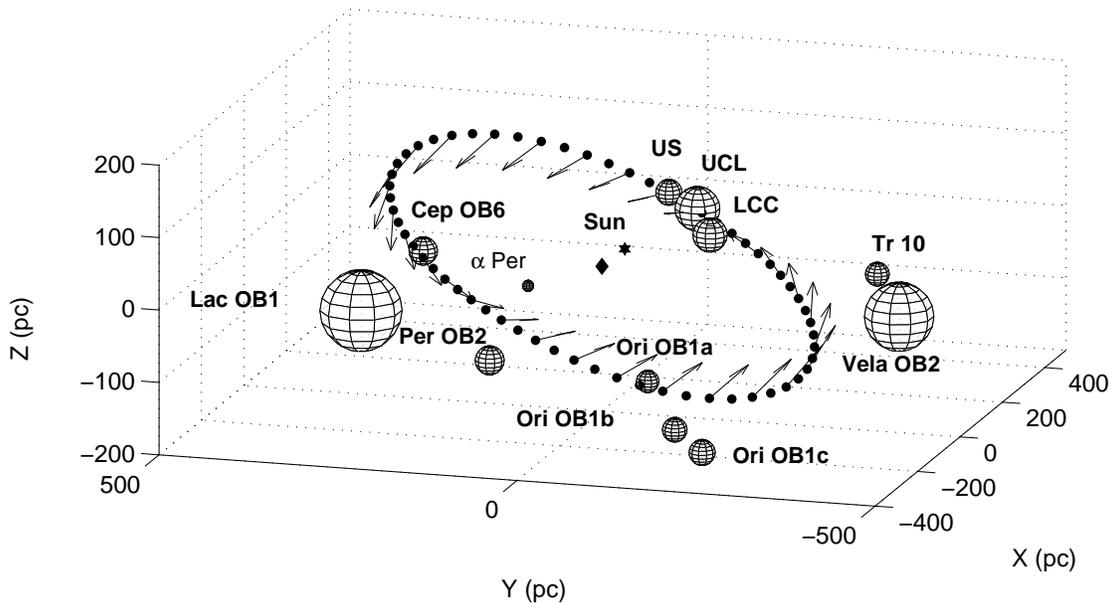,height=8cm}} \caption{3D view of the present Gould Belt and its
velocity field with respect to the local standard of rest. The local OB associations are marked as spheres.
The diamond notes the location of the Belt centre and the star that of the Sun.} \label{fig:Belt3D}
\end{figure}

Figure \ref{fig:Belt3D} displays the current Belt geometry with respect to the OB associations, the positions
of which are known from \textit{Hipparcos} measurements \cite{zeeuw99}. It nicely coincides with most of
them, but for Col 121 and Cep OB2 that clearly lie outside the Belt, and for Lac OB1 that is too far on the
wrong side of the Galactic plane. The total swept-up mass in the evolved cylindrical shell amounts to 2.4
10$^{5}$ M$\bigodot$. The mass accumulated over the past few Myr varies with longitude and is lowest in the
$35^{\circ} < l < 100^{\circ}$ and $-155^{\circ} <l < -110^{\circ}$ sectors which are indeed free of major
cloud complexes and OB associations near the rim. There is, however, no convincing relation between the mean
velocity field of the individual OB associations and the shell expansion \cite{perrot03}.

The column-densities of active X-ray stars show a marked excess over the Galactic population that varies with
longitude \cite{guillout98b}. It extends to the 300 pc visibility horizon in the $195^{\circ}-285^{\circ}$
interval, and to the remote edge at 170 pc of the Sco-Cen associations in the $285^{\circ}-15^{\circ}$
quadrant. It disappears in the $15^{\circ}-105^{\circ}$ interval, and is hardly visible out to 150 pc in the
$105^{\circ}-195^{\circ}$ quadrant, once the very nearby Hyades and Pleiades groups are subtracted. The lack
of active star formation within 50 pc is due to the Local Bubble. These distributions indicate that stellar
formation is not only active along the Belt rim, but also 100 pc or so inward, i.e. over a significant, yet
poorly constrained, radial extent. This picture is consistent with the spatial distribution of massive stars
in the 3rd and 4th Galactic quadrants. Yet, as for the young X-ray stars, no ring-like enhancement is seen in
the 1st and 2nd quadrants even though the OB star visibility reaches quite beyond the Belt edge.

\section{The Belt age and origin}
The Belt dynamical ages agree reasonably well for the different types of outburts (26.4 $\pm$ 0.4 Myr for the
3D cylindrical shock wave \cite{perrot03}, 23 and 15.5 Myr for a superbubble expansion with and without
internal pressure \cite{moreno99}), but they are notably smaller than that derived from the stars (30-60 Myr
for photometric ages \cite{comeron94a,torra00a}, 30-80 Myr from X-ray activity \cite{guillout98b}, and $34
\pm 3$ Myr from stellar dynamics \cite{comeron99}). The former are sensitive to the local gas density
estimates, the latter to the separation of the Belt and Galactic populations and to the photometric age
accuracy. To reduce this discrepancy, one may explore the influence of stellar rotation which can lower age
estimates by 30 to 50 \%, particularly for high rotators such as OB stars \cite{figueras98}.

The required initial kinetic energy of $(1.0 \pm 0.1)$ $10^{45}$ J in the 3D cylindrical shock wave
\cite{perrot03} is comparable, but 60\% higher than that needed in the 2D model \cite{olano82} because of the
extra work used to expand against the Galactic gravitational pull in the early phases. An equivalent energy
of 6 10$^{44}$ J is required for a superbubble expansion in a 3 times lower interstellar density, but against
ambient pressure \cite{moreno99}. This energy deposit is typical of multiple, rapidly succeeding supernovae
inside a young stellar cluster. It is also typical of a hypernova powering a collimated $\gamma$-ray burst
event, or of the potential energy of high-velocity clouds falling on the Galactic disc
\cite{comeron92,comeron94b}.

Preserving the structural coherence of the Belt stellar system over a large fraction of the vertical
oscillation and expansion timescales is challenging. Pure expansion models cannot reproduce the stellar data.
A system initially rotating as a solid body can remain flat and tilted in the Galactic gravitational
potential if it is initially inclined \cite{comeron99}. The angular momentum of the parent cloud would
therefore not be perpendicular to the Galactic plane. The dissolution of such a self-gravitating rotating
stellar system may explain the persistence of a flat expanding disc \cite{lindblad00}.

The Belt formation is still a puzzle. Its flatness and tilt remain very difficult to interpret. Various
scenarii involve the oblique impact of a high-velocity cloud on the Galactic disc \cite{comeron92,comeron94b}
or a cascade of supernova explosions (see \cite{poppel97} for a review). The former naturally provides some
inclination, the gas expansion, and it is consistent with the measured Oort constants. For instance
\cite{comeron92,comeron94b}, a 500 pc size, $10^{-2}$ M$\bigodot$ pc$^{-3}$ cloud, falling at 100 km/s from
the northern halo and from inside the solar circle, could have created the Gould Belt and the Monoceros R2
complex. Recent MHD simulations of an oblique impact, however, show that the hole punched in the Galactic
disc and the lateral compression waves are strongly driven by the vertical density gradient, perpendicular to
the Galactic plane, so getting a global inclination for the star-forming disc should be carefully
investigated (J. Franco, private communication). On the other hand, the expanding shock wave from an
explosive event or a rapid series of explosions \cite{olano82,perrot03} can match the space and velocity
distribution of the gas if an initial asymmetry provides a large inclination. Whether a $\gamma$-ray burst
event would apply is being investigated. Whether subsequent supernovae or stellar winds inside the Belt keep
powering its expansion at later stages is an open question. Injecting energy gradually would accelerate the
Belt expansion and further increase the age discrepancy ($R \propto t^{3/4}$ instead of $R \propto t^{1/3}$
in a uniform medium \cite{perrot03}), as well as reduce its eccentricity. The stellar orbits emerging from an
expanding superbubble \cite{moreno99} can reproduce the velocity field of the nearby O-B5.5 stars and of the
HI Lindblad ring only if the stars from the Pleiades group are removed, suggesting that two independent
events have formed the Pleiades and the Belt. Alternatively, a 2 10$^7$ M$\bigodot$, 400 pc size supercloud
has been proposed as the common precursor of the Sirius supercluster, the Gould Belt and the Local Arm
\cite{olano01}. The braking and compression of the supercloud while entering a spiral arm would have produced
the latter two while the stellar cluster, unaffected by friction, would have moved on, away from the gas
system. The supercloud angular momentum being concentrated at large radii, the inner regions would collapse
into a flattened disc, precursor of the Gould Belt, whereas the ejection of the outer parts into a super-ring
would form a precursor of the Local Arm.

Whatever triggered the Belt and its expansion has so deeply influenced the local interstellar medium, in
particular the pressure gradient, that the Local Bubble cavity, or rather the Local Chimney, has opened up to
the halo along an axis perpendicular to the Belt disc \cite{lallement03}. The chimney also notably opens
toward more intermediate- and high-velocity clouds than other regions of the halo. It has been suggested that
these clouds formed from material ejected by the initial Belt burst and they now fall back onto the Galactic
disc \cite{olano82,white93}. The lack of cold HI and the presence of intermediate-velocity clouds at high
latitude in the 2nd quadrant could be the signature of an explosive event that took place 35 myr ago near the
$\alpha$ Per association \cite{poppel00}.

\section{The Belt supernovae and cosmic rays}
During its evolution, the Belt has produced massive stars, therefore supernovae, in excess of the local
Galactic rate. Explosions should have lately occurred from the first generations of massive stars born in the
Belt. In the next few tens of Myr, 340 $\pm$ 30 stars \cite{cabrera99} with masses $>$ 8 M$_{\odot}$ will
explode and their maximum lifetime implies a crude minimum rate $>$ 35 collapses Myr$^{-1}$ kpc$^{-2}$. In
comparison, a Galactic rate of 20 events Myr$^{-1}$ kpc$^{-2}$ is inferred at the solar circle from the
distribution of stars in the Galaxy and from the average frequency of 2.5$^{+0.8}_{-0.4}$ events per century
for all types of supernovae in the Galaxy \cite{tammann94}, $\sim 85$ \% of which arise from the core
collapse of a massive star. This value reasonably agrees with the 29 progenitors Myr$^{-1}$ kpc$^{-2}$ found
with masses $>$ 8 M$_{\odot}$ within 1 kpc from the Sun \cite{tammann94}, in particular given the enhanced
yield from the Belt inside this region. A rate can be inferred for the recent past from the current Belt
stellar content as a function of mass, given the $\Gamma$ index of the initial mass spectrum ($dN/dM \propto
M^{\Gamma-1}$), lifetime estimates for stars with solar metallicity, a constant birth rate for simplicity, a
conservative mass threshold for collapse of 8 M$_{\odot}$, and a Belt age of 40 Myr consistent with the
stellar estimates. The observed star counts \cite{comeron94a} imply a supernova frequency of 20 to 27
supernovae per Myr in the entire Belt \cite{grenier00}. This rate falls to 17 to 20 supernovae per Myr using
revised star counts \cite{cabrera99}. The uncertainty stems from that in the star counts and Belt age and,
mostly, from the large uncertainty in the $\Gamma$ index between $-2.0$ and $-1.1$ at large mass. Using a
size of $354 \times 232$ pc, the corresponding rate of 65 to 78 Myr$^{-1}$ kpc$^{-2}$ is 3 to 4 times the
local Galactic one and is valid for the past few Myr. This rate stresses how actively the local medium has
been heated and enriched by supernova remnants as well as irradiated by cosmic rays.

The rate is consistent with the existence of four 0.1--1 Myr old radio loops \cite{berkhuijsen73}, the Local
Bubble, and possibly the Vela supernova remnant near the rim. Vela Junior, alias RX J0852.0-4622 or
G266.2-1.2, may be as close as 200 pc to be young enough to power the possible COMPTEL
detection\cite{tsunemi00} of $^{44}Ti$ decay lines. The interstellar absorption of the X-ray emission,
however, suggests a distance \cite{slane01} of 1--2 kpc, so a confirmation of the 68, 78, and 1157 keV lines
by INTEGRAL is eagerly awaited. RCW 114, alias G343.0-6.0, may be a nearby evolved remnant, well into its
radiative phase, that has expanded in a rather dense medium. Maps of the interstellar density do show a
cavity 200 pc away in this direction, near the Belt rim \cite{lallement03}.

Supernova shock waves are generally proposed as the sources of cosmic rays up to $10^{15}$ eV (see
\cite{drury01} for a review). The detection of synchrotron X-rays from various remnants (Cas A, SN 1006,
G347.3-0.5, Tycho) lends further support to the diffusive acceleration of electrons up to tens of TeV. Direct
observational evidence for the acceleration of ions is still searched for, but the indirect evidence becomes
compelling \cite{berezhko04}. The Belt rate is globally consistent with the power of 2.3 $10^{44}$ J
Myr$^{-1}$ kpc$^{-2}$ required to maintain the local cosmic-ray density \cite{blandford80} for a standard
supernova-to-cosmic-ray energy conversion efficiency of a few percent \cite{drury01}. However, the cosmic-ray
spectrum should vary with source proximity in space and time. The fluctuations are particularly strong for
the electrons above 50 GeV for which the synchrotron and inverse Compton radiative timescale is short, so
they diffuse no further than several hundred parsecs from their accelerator. Adding the contributions from
sources in the Galactic disc and in the Belt, with the respective rates quoted above, the average electrum
spectrum at Earth turns to be only slightly harder with the Belt than without \cite{pohl03}, with a spectral
index increase of 0.07 above 50 GeV. Yet, the currently measured spectrum may not be representative of the
average, nor should it be uniform in the local interstellar medium. It should correlate with that in the
nearby Ophiuchus and Taurus clouds, but not much with the spectrum in more remote places like Orion, Cepheus,
Perseus, and Monoceros \cite{pohl03}. The power per supernova required to sustain the local electron spectrum
is reduced by 40 \% when including the Belt, compared to a pure Galactic disc production. Cosmic-ray protons
and primary nuclei do not suffer serious radiative losses, but the source clustering in the Belt also leads
to an increased cosmic-ray density locally, to large fluctuations about the average, and to a slight
softening of the average spectrum because the higher-energy particles diffuse away faster \cite{bushing04}.

\begin{figure}
\centerline{\psfig{figure=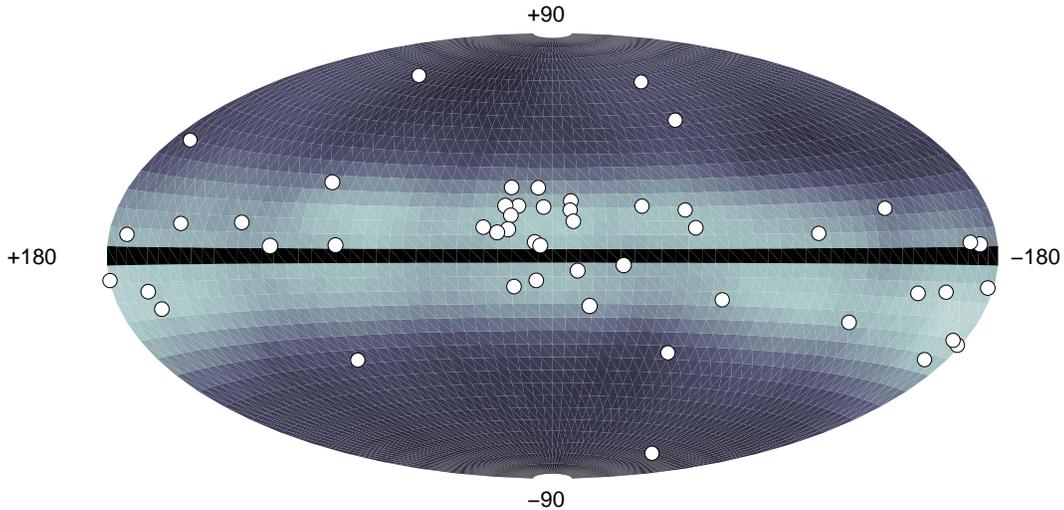,height=7cm}} \caption{Spatial distribution, in Galactic coordinates,
of the 50 unidentified persistent EGRET sources $|b| > 3^{\circ}$. The grey scale traces the column-density
of local young stars ($<$ B3), convolved with the EGRET exposure.} \label{fig:Beltufos}
\end{figure}

\section{$\gamma$-ray sources and pulsars in the Belt}
New facets of the Gould Belt activity have been brought to light at high energy with the discovery of a
population of $\gamma$-ray sources associated with it \cite{grenier00,gehrels00}. These sources are part of
the $\sim$ 130 unidentified $\gamma$-ray sources seen above 100 MeV by the EGRET telescope on board the late
Compton Gamma-Ray Observatory \cite{hartman99}. A subset of 35 to 40 of these sources \cite{grenier04}, among
the steadiest, is gathering at medium latitude ($3^{\circ} < |b| < 30^{\circ}$) along the characteristic
trace of the Gould Belt (see figure \ref{fig:Beltufos}). The source distribution is significantly better
correlated with the Belt than with other Galactic structures likely to display sources at medium latitude
\cite{grenier00}, such as a homogeneous spherical Galactic halo, or the local Galactic disc (with any scale
height), or the 400 pc-thick disc of radio pulsars. This should come as no surprise since there is clear
evidence, close to the Galactic plane, that unidentified $\gamma$-ray sources are linked to star-forming
sites. What is puzzling, however, is that most of the sources lack conspicuous radio or X-ray counterparts
despite their proximity. Whatever powers the $\gamma$-ray source emits most of its radiative energy in
$\gamma$ rays. No clear picture has emerged yet as to the nature of these objects, but neutron star activity
appears as a promising prospect. Other possibilities are unlikely \cite{grenier00}. The sources are too
bright in $\gamma$ rays to be unresolved gas clumps irradiated by the local cosmic-ray flux. The required
mass of $\sim 10^{4}$ M$_{\odot}$ at 500 pc cannot have escaped the radio and IR surveys, even when
considering radio beam dilution from unresolved clouds. Nor can they be slow, old neutron stars, wandering in
the interstellar medium and accreting gas from a dense cloud for they would be 10$^{2-3}$ times too rare and
the maximum Bondi-Hoyle accretion power of $\sim 2$ $10^{25}$ W that can be drawn from the surrounding HII
region is 10 times too low. The accretion power reached for slowly spinning, highly magnetized neutron stars
moving at 200--400 km s$^{-1}$ in the intercloud medium ($10^{-3}$ H cm$^{-3}$), though increased by
Kelvin-Helmholtz instabilities in the shocked gas, is also orders of magnitude too low. Isolated accreting
black holes are even more rare and compact objects accreting from a stellar companion, as well as
microquasars, would shine too brightly in X rays \cite{romero04}. No source coincides with any of the
numerous O Belt stars despite their highly supersonic winds with kinetic powers of $10^{28-29}$ W. Nearby
supernova remnants would appear as extended $\gamma$-ray sources. Therefore, only pulsars are left as
promising candidates even though the present models for pulsed emission predict too few of them.

13 radio pulsars from the ATNF catalogue are found with distances $<$ 1.5 kpc and age $<$ 2 Myr, but they are
too few and too fast to show a correlation with the inclined Belt or with the Galactic plane. The narrow
radio beams from many more may miss the Earth. Two young and energetic radio pulsars, Geminga and PSR
B0656+14, are born inside the Belt, 0.35 and 0.11 Myr ago, respectively. The younger Vela pulsar exploded 11
kyr ago near the rim. Geminga and Vela are both conspicuous $\gamma$-ray pulsars and the detection of the
third one above 100 MeV awaits confirmation. Geminga stands out as a unique example of a radio-quiet
$\gamma$-ray pulsar. It is the intrinsically faintest $\gamma$-ray pulsar observed so far, but equally faint
sources would have been easily detected anywhere inside or around the Belt if their radiation beam swept by
us. How the luminosity evolves at older ages is model dependent. Population synthesis studies have been
performed, both to provide information on the likelihood that these unidentified sources be radio-quiet
$\gamma$-ray pulsars and to study the pulsar population born in the Belt.

Pulsars are produced in the simulations with a constant birth rate of 1--1.4 per century over the past
billion years in the Galactic disc and of 20--24 per Myr in the expanding Gould Belt. They evolve in the
Galactic potential to the present time. Their initial period and magnetic field distributions are chosen to
fit the $P-\dot{P}$ diagram of a thousand radio pulsars. Their radio and $\gamma$-ray beams evolve with their
spin-down luminosity, $\dot{E}_{sd}$, as they get old. The slot-gap and outer-gap models
\cite{gonthier04a,cheng04} for pulsed emission predict similar luminosity evolutions ($L_{\gamma} \propto
\dot{E}_{sd}^{0.5}$ and $L_{\gamma} \propto \dot{E}_{sd}^{0.38}$, respectively). The beam apertures greatly
differ in the two cases because of their different origins in the open magnetosphere. Radiation is produced
in the slot gap in funnel-like beams deep inside the magnetosphere, above the polar caps, whereas the
outer-gap fan-like beams originate near the light cylinder. It appears that the characteristic spatial
signature of the Belt is preserved over several Myr despite its expansion, the rapid pulsar migration
\cite{arzoumanian02}, and the blending with the Galactic pulsar population \cite{perrot02,harding04}.

The results \cite{gonthier04a,cheng04} show that 4 slot-gap and 5 outer-gap radio-loud $\gamma$-ray pulsars
of Belt origin should have been detected by EGRET, in reasonable agreement with the 2 or 3 detections. An
extra 5 or 15 radio-quiet ones should appear as unidentified sources. The 3 times larger prediction of the
outer gap model results from the much wider beam that can be seen at large angles. In both cases, Belt
pulsars remain detectable for EGRET up to 2 Myr of age, i. e. much longer than for the more distant pulsars
in the Galactic disc. With more detections with the future GLAST satellite, to be launched in 2007, the ratio
of radio-loud to radio-quiet $\gamma$-ray pulsars will be an important clue to discriminate between pulsar
models. Yet, both models fail to explain the number of sources associated with the Belt. The same conclusion
is reached using a different scheme that yields an upper limit to the pulsar contribution at mid latitudes
for a minimal choice of assumptions, namely that the beam geometry shrinks with age as the open magnetosphere
and that the $\gamma$-ray luminosity scales with the spin-down power as for the known $\gamma$-ray pulsars
\cite{harding04}. A total of 19 Belt $\gamma$-ray pulsars is obtained by fitting the spatial and flux
distributions of all the unidentified EGRET sources near and away from the Galactic plane, without any radio
population constraints. So, the origin of most of the unidentified Belt sources remains a puzzle.

Finding the Belt neutron stars would provide a unique opportunity to constrain pulsar models to older ages
and a variety of aspect angles, and to study the progenitor mass threshold for producing a neutron star. As
supernova relics, they would bring valuable insight to the cosmic-ray production efficiency, the abundance of
explosive nucleosynthesis products, and to the filling factor of hot interstellar gas and its connection to
the halo. In other words, the Gould Belt is a lively and fascinating, but complex and out of the ordinary
place to live in and explore.

\end{document}